\documentclass[twocolumn,epjc3]{svjour3}  
\usepackage{rotating}
\usepackage{amsmath,amstext,amsbsy,amscd,bbm,epsfig,lscape}
\usepackage{graphicx,amsfonts,amssymb,color,comment,float,times}
\usepackage[utf8]{inputenc}
\definecolor{MyGrey}{rgb}{0,0,0} 
\definecolor{MyDarkBlue}{rgb}{0.,0.,1} 
\definecolor{MyLightBlue}{rgb}{0.22,0.51,0.9}
\usepackage{mathrsfs,amssymb,slashed}  
\usepackage{cancel}
\usepackage[normalem]{ulem}
\usepackage[colorlinks=true,linkcolor=MyDarkBlue,citecolor=MyDarkBlue,urlcolor=MyLightBlue,bookmarksnumbered=true,bookmarksopen]{hyperref}

\newcommand{\be}{\begin{equation}}
\newcommand{\ee}{\end{equation}}
\newcommand{\bea}{\begin{eqnarray}}
\newcommand{\eea}{\end{eqnarray}}
\newcommand{\GeV}{\,\text{GeV}}

\newcommand{\GeVsq}{\,\mathrm{GeV}^2}
\newcommand{\parenbar}[1]{\overset{
            \raisebox{-0.15em}{\scalebox{.4}{\textbf{(}}}
            \raisebox{-0.3em}{{\hspace{.03em}--\hspace{.05em}}}
            \raisebox{-0.15em}{\scalebox{.4}{\textbf{)}}}} {#1}}
\journalname{Eur. Phys. J. C}
\begin{document}

\title{Icecube/DeepCore tests for novel explanations of the MiniBooNE anomaly
\thanksref{t1}
}


\author{Pilar Coloma\thanksref{e1,addr1}
}

\thankstext{t1}{Preprint number: IFIC/19-27}
\thankstext{e1}{e-mail: pilar.coloma@ific.uv.es}

\institute{Instituto de F\'isica Corpuscular, Universitat de Val\`encia and CSIC, Edificio Institutos Investigaci\'on, Catedr\'atico Jos\'e Beltr\'an 2, 46980 Valencia, Spain \label{addr1}}

\date{Received: date / Accepted: date}

\maketitle

\begin{abstract}
While the low-energy excess observed at MiniBooNE remains unchallenged, it has become increasingly difficult to reconcile it with the results from other sterile neutrino searches and cosmology. Recently, it has been shown that non-minimal models with new particles in a hidden sector could provide a better fit to the data. As their main ingredients they require a GeV-scale $Z'$, kinetically mixed with the photon, and an unstable heavy neutrino with a mass in the 150~MeV range that mixes with the light neutrinos. In this letter we point out that atmospheric neutrino experiments (and, in particular, IceCube/DeepCore) could probe a significant fraction of the parameter space of such models by looking for an excess of ``double-bang'' events at low energies, as proposed in our previous work~\cite{Coloma:2017ppo}. Such a search would probe exactly the same production and decay mechanisms required to explain the anomaly.
\end{abstract}

\section{Introduction.} 
Since the discovery of neutrino oscillations, a tremendous effort in the neutrino sector has been made to measure the leptonic mixing parameters precisely and to test the three-neutrino oscillation paradigm. While most of the experimental results are perfectly consistent with oscillations in three families, several long-standing anomalies have sparked the interest of the neutrino community on light sterile neutrinos in the past two decades. 

The evidence favoring the existence of an eV sterile neutrino was first reported by LSND~\cite{Aguilar:2001ty}, an experiment designed to measure the oscillation probability in the \emph{appearance} channel $\bar\nu_\mu \to \bar\nu_e$. At the very small values of $L/E$ considered in LSND ($L$ being the distance to the detector, and $E$ the neutrino energy), standard neutrino oscillations have not yet started to develop and, therefore, a positive signal could indicate the existence of an extra sterile neutrino with a mass at the eV scale. The LSND result was later confirmed by the MiniBooNE experiment~\cite{AguilarArevalo:2010wv}. MiniBooNE used a higher energy neutrino beam and a longer baseline that at LSND but keeping the same value of $L/E$ in order to probe the same sterile neutrino mass scale. The collaboration reported excesses both in the $\nu_\mu \to \nu_e$ and $\bar\nu_\mu \to \bar\nu_e$ channels, with a significance which is now at the $4.7\sigma$~\cite{Aguilar-Arevalo:2018gpe}. 

However, in spite of this impressive statistical significance, the interpretation of the data in a 3+1 scenario (that is, adding a sterile neutrino to the three neutrinos in the Standard Model) suffers from tension on several fronts when confronted with the results from other experiments. First, the excess observed at MiniBooNE takes place at lower energies than expected from the LSND result. Moreover, a positive result in the appearance channels should be supported by a signal in the \emph{disappearance} channels ($\parenbar{\nu}_\mu \to \parenbar\nu_\mu$, and $\parenbar{\nu}_e \to \parenbar{\nu}_e$) as well, since in a minimal sterile neutrino scenario the probabilities in the appearance and disappearance channels are related. While reactor and radioactive neutrino experiments seem to observe a $\sim 3\sigma$ deficit of events with respect to theoretical predictions, all searches in the $\parenbar{\nu}_\mu \to \parenbar{\nu}_\mu$ disappearance experiments have been negative so far (for recent global fits to neutrino data see, e.g., Refs.~\cite{Giunti:2019aiy,Dentler:2018sju,Dentler:2017tkw,Collin:2016rao,Diaz:2019fwt}). This effectively rejects the minimal 3+1 sterile neutrio hypothesis at high confidence level~\cite{Giunti:2019aiy,Dentler:2018sju}. Finally, additional tensions arise from cosmological observables, both from measurements of the number of effective degrees of freedom at the time of Big-Bang Nucleosynthesis as well as from measurements of the sum of neutrino masses from the Cosmic Microwave Background and structure formation data~\cite{Ade:2015xua,Cyburt:2015mya}. 

In view of the difficulties that the vanilla 3+1 hypothesis is facing in order to explain the MiniBooNE low-energy excess (LEE), it is worth exploring non-minimal explanations. At this point, it is worth noting that while the MiniBooNE excess takes place in electromagnetic showers, the detector cannot distinguish if these are produced by photons or electrons. Therefore, an interesting possibility (to be tested by the MicroBooNE experiment at Fermilab~\cite{Acciarri:2016smi}) is that the excess comes from photons instead. Within the Standard Model (SM) framework, an excess of photons may come from cross section uncertainties or an underestimated background. While a promising candidate would be single-photon production in NC interactions~\cite{Hill:2009ek,Hill:2010zy,Zhang:2012xi,Wang:2013wva} it has been shown that this contribution is insufficient to successfully fit the observed excess~\cite{Wang:2014nat,Rosner:2015fwa}. Conversely, if the LEE is due to new physics (NP), it could be explained by the production of heavy neutrino that decays very promptly emitting a photon~\cite{Gninenko:2009ks,Gninenko:2010pr,Masip:2012ke}. Such scenarios successfully evade the constraints from disappearance experiments, as the heavy neutrino would not be produced in oscillations but in the up-scattering of light neutrinos inside the detector. Models of this type, where the neutrino has a non-standard transition magnetic moment, are able to fit the energy distribution observed at MiniBooNE better than a 3+1 hypothesis with an eV-scale sterile neutrino. However, if the production of the heavy state takes place through a photon this leads to a too forward-peaked angular distribution of events, which fails to reproduce the measurements reported by the collaboration. In addition, it is unclear that the required value of the transition magnetic moment is experimentally allowed by other constraints, see Refs.~\cite{McKeen:2010rx,Gninenko:2012rw}.

More recently, variations of the models presented in Refs. \cite{Gninenko:2009ks,Gninenko:2010pr,Masip:2012ke} have been put forward, where the heavy neutrino interacts with a new massive gauge boson ($Z'$) resulting from an extra $U(1)'$ symmetry~\cite{Ballett:2018ynz,Bertuzzo:2018itn}. The new particles introduced in this case interact with the SM fermions only via mixing: the $Z'$ is kinetically mixed with the photon, while the heavy neutrino mixes with the active neutrinos. In this case, the heavy neutrino would also be produced in up-scattering of light neutrinos, but in this case the interaction would be mediated by the $Z'$. Once produced, it would travel a macroscopic distance and decay via the new interaction, leading to an $e^+ e^-$ pair in the final state: if the two Cherenkov rings overlap, the observed signal would also be misidentified as a single electron-like event at MiniBooNE. Thanks to the newly introduced \emph{massive} gauge boson, the angular distributions obtained are less forward-peaked, allowing the model to successfully fit \emph{both} the energy and angular distributions observed at MiniBooNE. Moreover, besides explaining the LEE, these scenarios are also theoretically well-motivated: a minimal modification of the phenomenological models proposed in Refs.~\cite{Ballett:2018ynz,Bertuzzo:2018ftf} with two hidden states may be able to generate the SM neutrino masses \cite{Bertuzzo:2018ftf,Ballett:2019cqp} and could even accommodate a dark matter candidate~\cite{Ballett:2019pyw,Blennow:2019fhy}.

Given that the interactions to the SM fermions are heavily suppressed via mixing, the parameter space where the dark neutrino models are able to explain the MiniBooNE anomaly is difficult to probe experimentally. Modifications to the $\nu_\mu$ neutral-current (NC) scattering cross section with nucleons would take place only at the percent level, well below current experimental uncertainties~\cite{Acero:2019qcr} (for reviews, see e.g. Refs.~\cite{Formaggio:2013kya,Alvarez-Ruso:2017oui}). However, it has been recently pointed out that relevant constraints can be derived from $\nu-e$ scattering data, given the electron-like nautre of the produced signal from the decay of the heavy neutrino~\cite{Arguelles:2018mtc}. In fact, the authors of Ref.~\cite{Arguelles:2018mtc} have reanalyzed CHARM-II and MINERvA data and their results disfavor the model proposed in Ref.~\cite{Bertuzzo:2018ftf} in the region of parameter space where both the MiniBooNE energy and angular distributions are successfully reproduced. However, while for the model in Ref.~\cite{Bertuzzo:2018ftf} the heavy neutrino production cross section would be coherent, for the model from Ref.~\cite{Ballett:2018ynz} the incoherent contribution would be dominant (due to the higher $Z'$ mass considered), and a separate analysis would be required to derive a constraint.

In this Letter we point out that atmospheric neutrino experiments could probe a significant fraction of the allowed parameter space of this class of models and, in particular, for the model proposed in Ref.~\cite{Ballett:2018ynz}. For large values of the heavy neutrino mixing we find that the model would lead to a significant excess of NC-like events, since the NP cross section would be comparable to the SM NC cross section. In addition, the heavy neutrino would be relatively long-lived and could propagate over macroscopic distances in the detector after being produced. As it decays, it may lead to a separate second shower inside Icecube/DeepCore. Thus, for small values of the active-sterile neutrino mixing a search for ``double-bang'' (DB) events at low energies, as proposed in our previous work~\cite{Coloma:2017ppo}, would lead to impressive sensitivities. Such a search would probe exactly the same production and decay mechanisms needed to explain the anomaly. 


\section{ Model details.} The model proposed in Refs.~\cite{Ballett:2018ynz,Bertuzzo:2018itn} extends the SM gauge group with an additional $U(1)'$ symmetry, which is however broken at low energies. A priori, it is assumed that none of the SM fermions are charged under the new symmetry. However, unless explicitly forbidden, the new gauge boson $X_\mu$ associated to the hidden $U(1)'$ symmetry will kinetically mix with the SM hypercharge boson through a term of the form~\cite{Holdom:1985ag} $B_{\mu\nu}X^{\mu\nu}$, where $B$ and $X$ stand for the hypercharge and $U(1)'$ field strength tensors. This induces couplings to the SM fermions that are suppressed by the kinetic mixing parameter $\chi$. At first order in $\chi$, the $Z'$ will interact with the SM fermions through the following term in the Lagrangian:
\begin{equation}
\mathcal{L} \supset -e q_f c_W \chi \bar{f} \gamma^\mu f Z_\mu'   \, ,
\end{equation}
where $Z'_\mu$ is the mass state associated to $X_\mu$, $c_W$ is the cosine of the weak mixing angle, $e$ is the electron charge and $q_f$ is the charge of the fermion $f$. The model also requires the addition of a fourth massive neutrino which does not couple directly to any of the SM gauge bosons but couples to the $Z'$ directly. The whole neutrino flavor and mass bases are related by the usual unitary transformation $\nu_\alpha = \sum_i U_{\alpha i} \nu_i$, where $i =1,2,3,4$ and $\alpha=e,\mu,\tau,s$ refer to the mass and flavor indices, respectively. As a result, the active neutrinos will also inherit interactions with the $Z'$ which are suppressed with the mixing to the sterile state:
\begin{eqnarray}
\mathcal{L} & \supset & U^*_{\alpha 4} g' \bar\nu_\alpha \gamma^\mu P_L \nu_4 Z'_\mu + \nonumber \\
& & U^*_{\alpha 4} U_{\beta 4} g' \bar\nu_\alpha \gamma^\mu P_L \nu_\beta Z'_\mu + 
g' \bar\nu_4 \gamma^\mu P_L \nu_4 Z'_\mu  \, ,
\end{eqnarray}
where $g'$ is the coupling constant between the fourth neutrino and the $Z'$. At this point, it should be stressed that the mixing between active neutrinos and sterile neutrinos in the MeV-GeV range is tightly constrained in the $\mu$ and $e$ sectors thanks to precision measurements of $\beta$- and meson decays in the laboratory. In the $\tau$ sector, on the other hand, laboratory searches are much weaker due to the intrinsic difficulties of producing mesons with large branching ratios into $\nu_\tau$. In the region of interest ($m_4 \sim 150$~MeV) the most relevant constraints come from CHARM~\cite{Orloff:2002de} and NOMAD~\cite{Astier:2001ck} data (see e.g., Refs.~\cite{Atre:2009rg} and~\cite{Drewes:2015iva} for a compilation of bounds). However, since in the model considered here the heavy neutrino would decay very promptly these bounds would be considerably relaxed, and values as large as $|U_{\tau 4}|^2 \sim 10^{-3}$ are still allowed~\cite{Ballett:2018ynz}. Finally, while astrophysical constraints may in principle be relevant for this type of models (see e.g. Ref.~\cite{Rembiasz:2018lok}), these are successfully avoided in the region of parameter space considered in Ref.~\cite{Ballett:2018ynz}.

At MiniBooNE an intense neutrino flux is produced from meson decays, resulting in a neutrino energy distribution that peaks at around 500~MeV. The beam composition is primarily $\nu_\mu$ in neutrino mode, and $\bar\nu_\mu$ in anti-neutrino mode. The requirements that the model should satisfy in order to fit the MiniBooNE LEE are summarized as follows (we refer the interested reader to Ref.~\cite{Ballett:2018ynz} for further details):
\begin{enumerate}
\itemsep0em
\item The NP should induce an up-scattering cross section for $\nu_\mu$ to the heavy state $ \sigma_{\mu}^{Z'} \sim 0.01 \sigma_{SM}$ (in the QE regime), where $\sigma_{SM}$ is the SM NC cross section. This effectively imposes a constraint on the combination $g'^2 \chi^2 |U_{\mu 4}|^2/m_{Z'}^4$, and favors low masses for the $Z'$ (at the GeV scale) in order to reach large enough cross sections. 
\item The masses of the new particles introduced in the model control the energy and angular distributions of the events. In particular, the neutrino mass should lie between 100 and 200 MeV in order to be able to fit the observed energy distribution. The angular distribution, on the other hand, is sensitive to the $Z'$ mass. For the model considered in Ref.~\cite{Ballett:2018ynz} (where $m_{Z'} > m_4$), $Z'$ masses below a GeV are disfavored by the fit. 
\item The decay of the neutrino should take place inside the MiniBooNE detector, which imposes a requirement on its decay length in the lab frame of $L_{lab}^{mboone}\lesssim \mathcal{O}(1)~\mathrm{m}$. For a heavy neutrino with a mass $m_4\sim 150~\mathrm{MeV}$ and energies around 500~MeV, this implies a lifetime $c\tau \sim 0.3$~m. This is satisfied setting $|U_{\tau 4}|^2 \sim 8\times 10^{-4}$. 
\item In addition, the branching ratio of the heavy neutrino into the channel $\nu_4 \to \nu_\alpha e^+ e^-$ should be dominant. Since $ |U_{\tau 4}|^2 \gg |U_{\mu 4}|^2, |U_{e4}|^2$, the decay takes place predominantly into $\nu_\tau e^+ e^-$. 
\end{enumerate}


\section{Heavy neutrino production in atmospheric neutrino experiments.} Atmospheric neutrinos are a by-product of meson decays produced when cosmic rays hit the top layers of the atmosphere. While the resulting flux is primarily composed by $\nu_\mu$ and $\bar\nu_\mu$ (resulting from pion and kaon decays), standard neutrino oscillations in the $\nu_\mu \to\nu_\tau$ channel provide a sizable $\nu_\tau$ flux contribution for neutrino trajectories crossing the Earth. 

To get an estimate on the expected number of heavy neutrinos produced in atmospheric neutrino detectors, it is useful to compare the expected size of the NP cross section to the NC cross section in the SM. Within the parton model, the double differential cross section for the up-scattering in a neutrino-nucleon interaction reads:
\begin{equation}
\label{eq:xsec}
\frac{d^2\sigma_\alpha^{Z'}}{dx dy} = \frac{2 G_\alpha'^2 M E_\nu }{\pi}
\frac{M_{Z'}^4 (q_u^2 + q_d^2)}{(Q^2  + M_{Z'}^2)^2}  
[1 + (1-y)^2 ] \mathcal{F}(x) \, ,
\end{equation}
where $M$ is the proton mass, $E_\nu$ is the initial neutrino energy, $x$ is the fraction of the nucleon momentum carried by the parton and $y\equiv\nu/E_\nu$, with $\nu=E_\nu - E_N$ being the energy transferred in the interaction and $E_N$ the energy of the heavy neutrino in the final state. Here, $\mathcal{F}(x) = x\sum_{q}(f_q(x) + f_{\bar q}(x))$ contains the dependence on the parton distribution functions (PDFs) of the proton ($f_{q,\bar q}$). In deriving Eq.~\eqref{eq:xsec}, any effects due to the mass of the heavy neutrino in the final state have been neglected (since we are mainly interested in the mass range below 200~MeV), and an isoscalar target nucleus has been assumed. Moreover, we have introduced an effective coupling constant $G_\alpha'$ (analogous to the Fermi constant in the SM, $G_F$) which depends on the flavor of the incident light neutrino, $\alpha$:
\begin{equation}
\label{eq:Galpha}
\frac{G_\alpha ' }{\sqrt{2}} = \frac{g' U_{\alpha 4} \chi e c_w}{2 M_{Z'}^2} \, .
\end{equation}
As seen from Eq.~\eqref{eq:xsec}, the comparison to the SM cross section cannot be performed in a straightforward manner due to the different chirality of the currents involved (left-handed in the SM, as opposed to vector currents in the $Z'$ case). However, considering that the final event sample contains contributions from both neutrino and antineutrino fluxes (with similar intensities) interacting both on protons and neutrons (for an isoscalar target), the number of heavy neutrinos produced $N_{\alpha}^{Z'}$ due to the up-scattering process $\nu_\alpha \mathcal{N} \to \nu_4 \mathcal{X}$ (where $\mathcal{N}$ is a nucleon and $\mathcal{X}$ is the final state hadron(s)) can be taken as approximately proportional to the number of $\nu_\alpha$-nucleon NC events in the SM, $N_{\alpha}^{Z'} \simeq \epsilon_\alpha N^{Z}_{\alpha}$. The proportionality constant reads
\begin{equation}
\epsilon_\alpha \simeq  \frac{G_\alpha'^2 }{G_F^2 } \frac{ 2 (q_u^2 + q_d^2)}{ 
g_{L,\nu}^2 (g_{L,u}^2  + g_{L,d}^2 ) } \left( \frac{M_{Z'}^2}{\langle Q^2\rangle  + M_{Z'}^2} \right)^2\, ,
\label{eq:epsilon}
\end{equation}
with $g_{L,f} = T_{3,f} - q_f\sin^2\theta_w$. Here, $T_{3,f}$ is the weak isospin of fermion $f$ while $\langle Q^2 \rangle$ should be taken as the typical value of the squared momentum transfer involved in the interaction, and we have neglected $g_{R,q}$ (since $g_{L,q}^2 \gg g_{R,q}^2 $). Note that the energy transfer $\nu$ and the invariant mass of the hadronic shower $W^2$ relate to the value of $Q^2$ as $ Q^2 = 2M\nu + M^2 - W^2$. 

An estimate on the number of events can be obtained for the benchmark values given in Ref.~\cite{Ballett:2018ynz} that provide a best-fit to the MiniBooNE anomaly: $|U_{\tau 4}|^2 = 7.8\times 10^{-4}$, $|U_{\mu 4}|^2 =1.5\times 10^{-6}$, $\chi^2=5\times 10^{-6}$, $g'=1$ and $ m_{Z'}=1.25$~GeV. For DIS, assuming $\langle Q^2 \rangle \sim \mathcal{O}(5)\GeVsq$, this leads to $\epsilon_\tau \sim 0.5$, and $\epsilon_\mu \sim 10^{-3}$ (or, more generally, $\epsilon_\alpha \sim 6\times 10^{2} |U_{\alpha 4}|^2 $).
From this naive estimate it is easy to see that a large number of $N_{\tau}^{Z'}$, indistinguishable from SM NC events, would be expected at atmospheric neutrino experiments due to the relatively large values of $U_{\tau 4}$ required in order to fit the LEE. A search for an excess of cascade events was proposed in Ref.~\cite{Masip:2014xna} to test the model from Ref.~\cite{Masip:2012ke}, and in fact a similar search could also be performed in this case. However, it should be noted that while in Ref.~\cite{Masip:2014xna} the excess would take place for down-going events, for the model considered in this work the excess would appear in the up-going sample. A search of this sort naively presents several advantages with respect to a DB search. On one hand, for the benchmark parameters considered the size of the signal may be much larger than for DB events. Moreover, while observing and identifying two separate showers may be challenging from the experimental point of view, the observation of NC-like events represents an easier task for most experiments. However, this signal would be subject to much larger backgrounds, coming not only from standard neutrino NC events but also from $\nu_e$ and $\nu_\tau$ CC interactions in the detector. 

Similarly, a search for an excess of NC-like events at the T2K experiment may also be able to constrain the minimal realization of the model from Ref.~\cite{Ballett:2018ynz}, since by the time the neutrino beam reaches the detector most of the muon neutrinos have already oscillated into tau neutrinos. A possibility to avoid this would be to considerably reduce the size of $|U_{\alpha 4}|^2$, in order to bring the number of events due to the NP below experimental uncertainties\footnote{It should be stressed that, in the minimal model considered in Ref.~\cite{Ballett:2018ynz}, this would lead to a value of $c\tau$ too large to explain the MiniBooNE LEE. Nevertheless it might be possible to work around this in non-minimal variations of the model. }. In this case, a DB search at Icecube/DeepCore would still yield impressive sensitivities, as we will see in the next sections.


\section{ Icecube/DeepCore Double-Bangs to test the anomaly. } The IceCube South Pole neutrino telescope contains over 5000 Digital Optical Modules (DOM) deployed between 1450 m and 2450 m below the ice surface. When high-energy charged particles travel through the ice, they emit radiation that is then detected by the DOMs. At Icecube, two event topologies are distinguished: \emph{track} events, produced by muons, and \emph{cascades} (or showers), produced by other charged particles such as electrons or hadrons. The inner core of Icecube, approximately 2100 m below the surface cap, is called DeepCore. With a DOM density roughly five times higher than that of the standard IceCube array, DeepCore can observe showers with much lower energies than Icecube (down to $E \sim 5.6$~GeV). 

At Icecube, DB events are a standard signal for $\tau$ leptons at ultra-high energies, where the boosted decay length of the $\tau$ is long enough to be able to resolve the two showers from its production and decay vertices. In this work, however, we will be considering much lower energies as proposed in our previous work~\cite{Coloma:2017ppo}. In this context, a DB event is defined as an event that satisfies the following conditions: (a) it should lead to two distinct showers separated by a macroscopic distance of at least 20~m; (b) each of the showers should have a minimum energy of 5.6~GeV in order to be observed at the DOMs; and (c) the first event should be observed a a minimum of three (or four) DOMs if the event takes place inside (outside) DeepCore, in order to set the detection trigger off. Allowing events to trigger the detector outside the DeepCore volume may lead to additional backgrounds from atmospheric muons. We expect a sufficient reduction of this background by requiring that the two DB events fall on a straight line and take place in the up-going direction, which should be achievable given the very good timing resolution of the DOMs~\cite{Aartsen:2015dlt}. At low energies and in the up-going direction, the most important background that could induce a DB would be coincidental showers from atmospheric neutrino events (estimated at 0.05 events/yr, see Ref.~\cite{Coloma:2017ppo} for details). However, a careful computation of the background levels for this search should be carried out within the experimental collaboration.

From a model-independent perspective, the DB event rate at Icecube depends solely on two physics observables: the value of the production cross section inside the detector, which determines the number of heavy neutrinos produced in the interaction of atmospheric neutrinos; and the decay length of the heavy neutrino in the lab frame, $L_{lab}$, which determines if the second decay will occur inside the detector. Of course, $L_{lab}$ eventually depends on the value of the heavy neutrino lifetime $c\tau$, its mass $m_N$ and its energy $E_N$. The total number of DB events per unit of time, where the production vertex takes place in the up-scattering of $\nu_\alpha$, can be expressed as:
\begin{eqnarray}
\frac{dN^{DB}_{\alpha}}{dt} &\! = \! \rho_{n} \sum_{\nu, \bar\nu} & \! \! \int \! dE_N dE_\nu d c_\theta \frac{d\phi_{\mu}(E_\nu, c_\theta) }{dE_\nu d c_\theta} P_{\mu\alpha}(E_\nu, c_\theta) \nonumber \\ 
& & 
\frac{d\sigma^{Z'}_{\alpha}(E_\nu, E_N)}{dE_N} V_{det}(L_{lab}, c_\theta) \, , \label{eq:dNdt}
\end{eqnarray}
where $\rho_{n}$ is the average nucleon density in the ice, $c_\theta$ is the cosine of the zenith angle $\theta$, $\phi_\mu$ is the atmospheric $\nu_\mu$ (or $\bar\nu_\mu$) flux, $\sigma_{\alpha}^{Z'}$ is the cross section for the production of the heavy (anti)neutrino in $\parenbar{\nu}_\alpha$-nucleon interactions,  $P_{\mu\alpha}$ is the standard oscillation probability in the $\parenbar{\nu}_\mu \to \parenbar{\nu}_\alpha$ channel, and $V_{det}$ is the effective volume of the detector, which depends on the heavy neutrino decay length in the lab frame and on the zenith angle. 

The effective volume of the detector was computed in Ref.~\cite{Coloma:2017ppo} via Monte Carlo integration, where it was found to be maximal for decay lengths in the lab frame of around $L_{lab} \sim 100$~m. For smaller decay lengths, the effective volume decreases since the density of the DOM grid is too low to be able to distinguish the two showers. For much longer decay lengths, on the other hand, it decreases roughly as $1/L_{lab}$ as the neutrinos will typically exit the detector before decaying. The effective volume also depends on the zenith angle, as expected from geometrical arguments, and is maximized for trajectories crossing the Earth's core ($\cos\theta \simeq -1$), where we expect the transition probability $\nu_\mu \to \nu_\tau$ to be maximal for energies around $E_\nu \sim 25~\GeV$. Coincidentally, at these energies, for $m_4=150$~MeV and $c\tau \sim 0.3~\mathrm{m}$ we find a boosted decay length of 50~m, precisely in the region where we expect the effective volume of the detector to be close to maximal. As a reference value, for $L_{lab}\sim 50~\mathrm{m}$ and $\cos\theta=-1$ we find $V_{det}\simeq 0.05~\mathrm{km}^3$. This is larger than the DeepCore volume, since we allow events to trigger the detector outside the DeepCore volume if at least four DOMs are hit simultaneously. We have checked that if we additionally require that the first event takes place inside DeepCore we recover its volume, as expected. 


\section{Numerical results.} Our exact numerical calculation agrees reasonably well with the naive estimate outlined above. 
The number of DB events have been computed following Eq.~\eqref{eq:dNdt}. The cross section has been computed as outlined in Eq.~\eqref{eq:xsec} with the CTEQ6.6 set of parton distribution functions~\cite{Nadolsky:2008zw,Buckley:2014ana}, together with the Parton package~\cite{partonpackage} for their numerical evaluation in Python3. For simplicity, only the contributions from $u, \bar u, d, \bar d$ quarks have been considered, which is expected to be a good approximation within the range of momentum transfer considered. In order to ensure that the two showers are above 5.6~GeV and that the interaction falls in the DIS regime, the following cuts are imposed: $5.6 \leq (\nu/\GeV) \leq E_\nu - 5.6$, $Q^2 \geq 4$~GeV$^2$, and $W^2 > 1.5$~GeV$^2$. Neutrino oscillation probabilities are computed using the GLoBES package~\cite{Huber:2007ji,Huber:2004ka} dividing the Earth matter density profile into 20 different layers, each of them with a constant density according to the PREM model~\cite{Dziewonski:1981xy,prem2}. The values of the oscillation parameters are taken at the best fit from a global analysis of neutrino oscillation data~\cite{Esteban:2018azc,nufitwebpage}. The atmospheric neutrino and antineutrino fluxes have been computed for zenith angles in the range $-1 \leq \cos\theta \leq 0$ using the MCEq module~\cite{Fedynitch:2015zma,Fedynitch:2012fs} for Python2, with the SYBILL-2.3 hadronic interaction model~\cite{Fedynitch:2018cbl}, the Hillas-Gaisser cosmic-ray model~\cite{Gaisser:2011cc} and the NRLMSISE-00 atmospheric model~\cite{Picone:2002go}. Finally, the effective volume of the detector is computed using Monte Carlo integration, as in Ref.~\cite{Coloma:2017ppo}.

\begin{figure*}[ht!]
\begin{center}
\begin{tabular}{cc}
\includegraphics[width=\columnwidth]{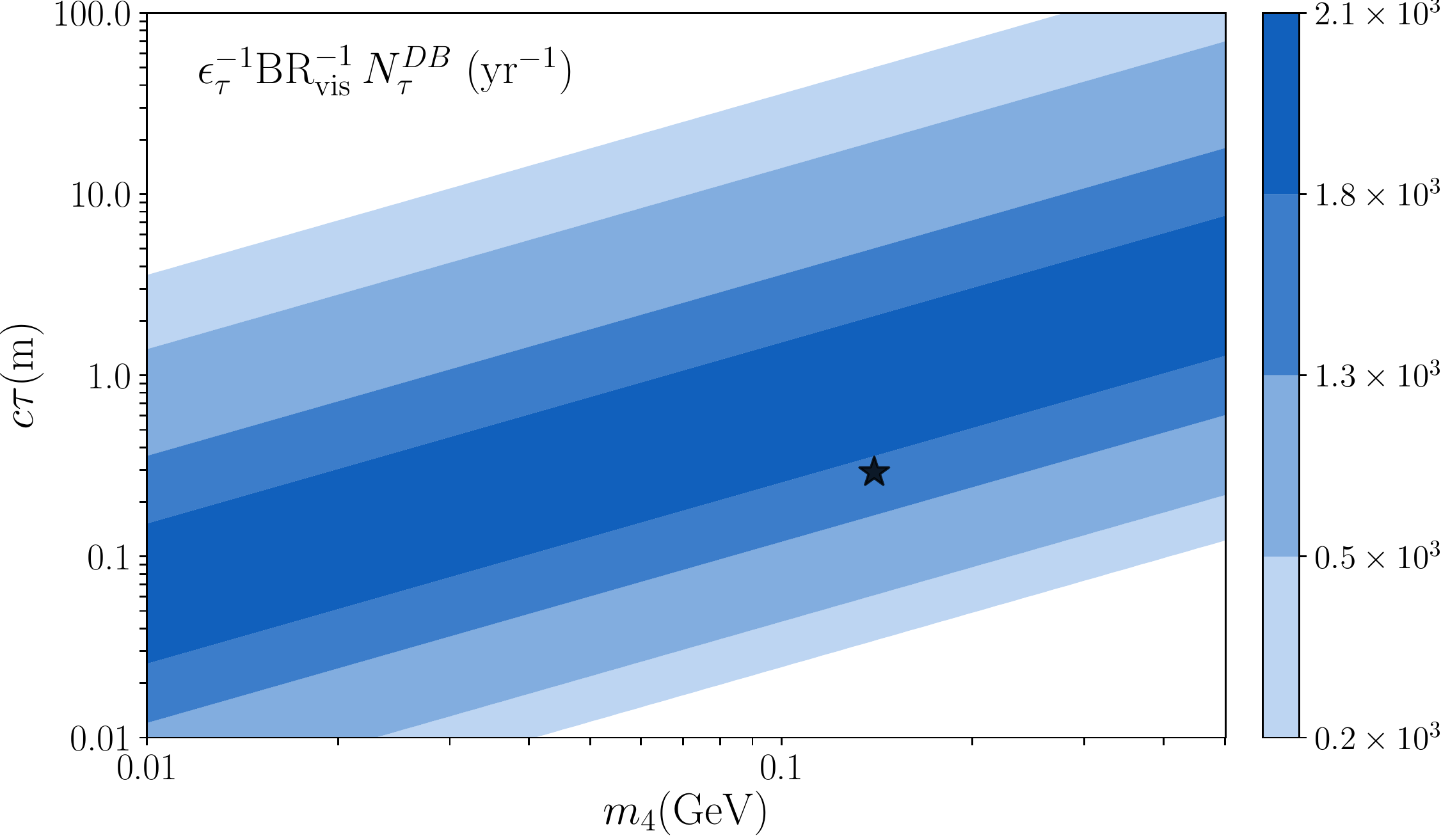} &
\includegraphics[width=\columnwidth]{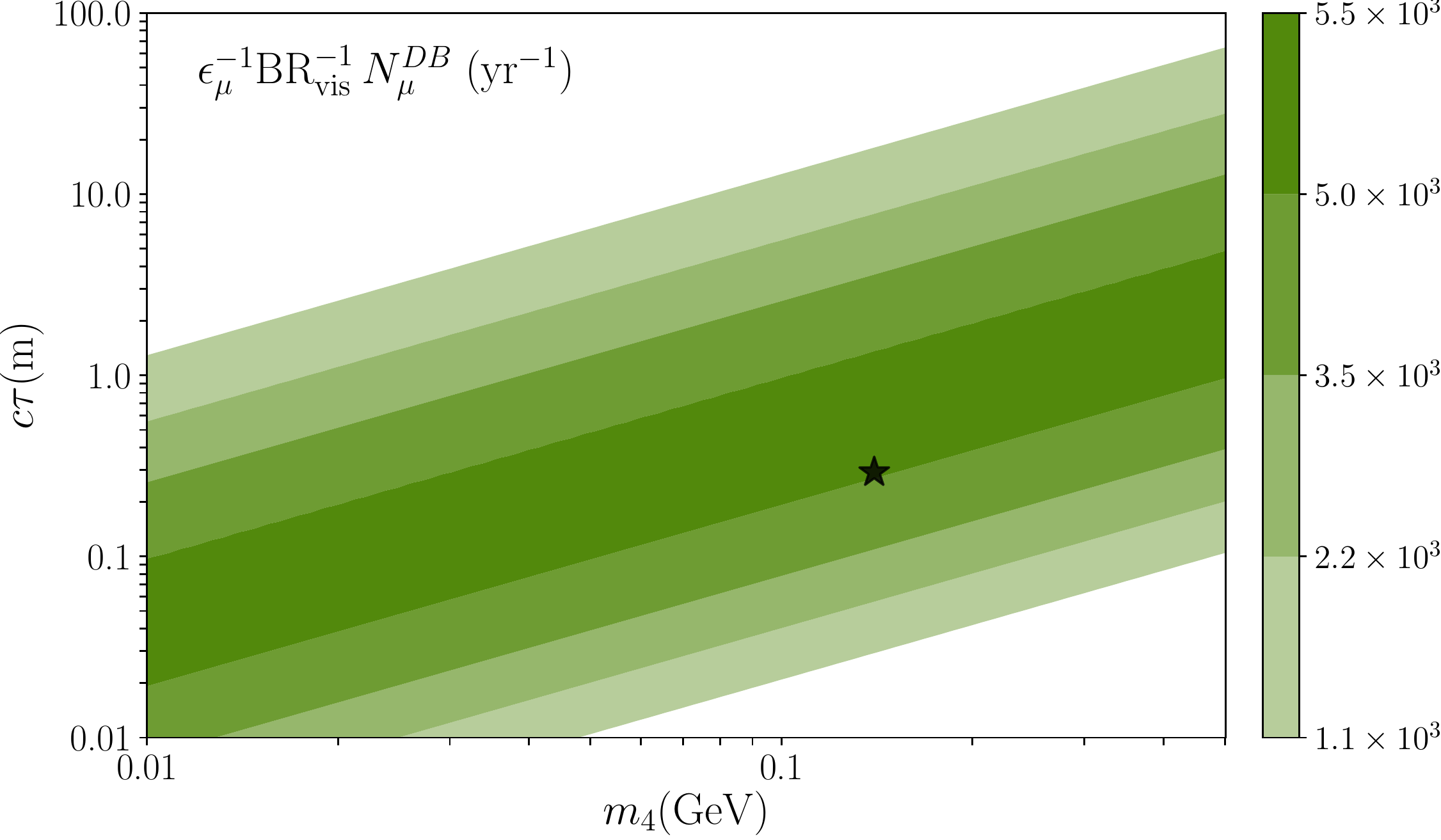}
\end{tabular}
\caption{ Expected number of DB events per year at Icecube/DeepCore. The left panel shows the result for $\nu_\tau$ up-scattering into the heavy neutrino, while the right panel considers $\nu_\mu$ up-scattering. The black star indicates a representative point in the parameter space for the model of Ref.~\cite{Ballett:2018ynz}, corresponding to $m_4=140$~MeV and $c\tau~\sim 0.3~\mathrm{m}$. Note that the benchmark values required to fit the MiniBooNE LEE imply $\epsilon_\tau \sim 0.5$ and $\epsilon_\mu \sim 10^{-3}$ (assuming $\langle Q^2\rangle \sim 5\GeVsq$, see Eq.~\eqref{eq:epsilon}).
\label{fig:ctau-m} }
\end{center}
\end{figure*}
Our main result is given in Fig.~\ref{fig:ctau-m}, where the colored bands indicate the regions of parameter space where the expected number of signal events per year would exceed a certain number, as indicated by the legend. While the number of events has been exactly computed following the procedure described above, our results are given as a function of $\epsilon_\alpha$ and the branching ratio of the neutrino decay into visible particles, $\mathrm{BR}_{vis}$. This way we ensure that our results can be easily adapted to non-minimal versions of this model simply multiplying by the corresponding value of the product $\epsilon_\alpha \mathrm{BR}_{vis}$. Note that while at MiniBooNE the up-scattering would take place exclusively in $\nu_\mu$ interactions, at Icecube/DeepCore both $\nu_\mu$ and $\nu_\tau$ fluxes are available and may up-scatter into $\nu_4$. However, since the expected $\nu_\mu$ and $\nu_\tau$ fluxes will be very different due to standard neutrino oscillations, our results are provided for the two cases separately. 

Given the low background level expected in the SM for this type of signal~\cite{Coloma:2017ppo}, even a handful of events may be enough to reject the background-only hypothesis at high confidence. Therefore, if the heavy neutrino is mixed with $\nu_\tau$, we find that a DB search may be sensitive to models with values of $\epsilon$ as low as $\epsilon_\alpha \sim \mathcal{O}(10^{-3})$, as long as the lifetime and mass of the heavy neutrino fall inside the region $5 \lesssim \frac{c\tau}{\mathrm{m}}\frac{\GeV}{m_4} \lesssim 10$. It should be noted that $\nu_\tau$ up-scattering takes place at $ E_\nu \sim 25$~GeV (where the oscillation probability $\nu_\mu \to\nu_\tau$ is close to 1), while for $\nu_\mu$ up-scattering the heavy neutrino production typically takes place at energies above 30~GeV (where the flux does not oscillate). Therefore, the higher neutrino energies in the $\nu_\mu$ case with respect to the $\nu_\tau$ case leads to maximal sensitivities in the right panel for lower values of $c\tau$. Finally, in Fig.~\ref{fig:ctau-m} the black star indicates the benchmark point from Ref.~\cite{Ballett:2018ynz}, which provides a best-fit to the MiniBooNE LEE. Assuming $\mathrm{BR}_{vis}\sim \mathcal{O}(1)$ we obtain, for their benchmark values, approximately $10^3$ DB events/yr for $\nu_\tau$ up-scattering and 6 DB events/yr for $\nu_\mu$ up-scattering. 


%
\section{Conclusions.}
In this Letter we have shown that the minimal model in Ref.~\cite{Ballett:2018ynz} proposed to explain the MiniBooNE anomaly would lead to a cross section for $\nu_\tau$ up-scattering into the heavy state comparable to the SM NC cross section. Consequently, this would yield a significant excess in NC-like events with respect to the SM prediction, observable in the up-going sample at atmospheric neutrino experiments. Modified versions of this model, however, may be able to avoid such large values of $U_{\tau 4}$. We have shown that in this case Icecube/DeepCore could search for an excess of DB events~\cite{Coloma:2017ppo}, for heavy neutrino production cross section as low as the per mille level of the SM NC cross section. However, while the assumptions made in our computation of the DIS cross section have been generally conservative, a careful study of the actual sensitivity of this signal is needed. This should eventually be carried out by the experimental collaboration, in order to include detection efficiencies and a proper Monte Carlo simulation of the expected background rates. 

As a final remark it is also worth pointing out that, even though this work is mainly motivated by the model proposed in Ref.~\cite{Ballett:2018ynz}, our result is more general. In particular, further modifications of the model could avoid the correlation between production and decay processes (either completely or partially), since this eventually depends on its particle content and on the mixing between the heavy and SM neutrinos. Following a completely model-independent approach, in our numerical analysis we have precisely assumed that the production and decay mechanisms may be completely decoupled. Of course, once a particular model is specified, the available region of parameter space will be limited to a subset of the full region shown in Fig.~\ref{fig:ctau-m}, depending on how $\sigma$, $m_4$ and $c\tau$ relate to one another.

\begin{acknowledgements}
The author is especially grateful to Pilar Hernandez for illuminating discussions, to Matheus Hostert for pointing out a mistake in the estimate of the number of events and to Enrique Fernandez-Martinez and Ian Shoemaker for a careful reading of the manuscript. She warmly thanks as well Carlos Arg\"uelles, Andrea Donini, Jacobo Lopez-Pavon, Ivan Martinez-Soler and Carlos Pena for useful discussions. This work has been supported by the Spanish MINECO Grants FPA2017-85985-P and SEV-2014-0398, and by the European Union’s Horizon 2020 research and innovation program under the Marie Sklodowska-Curie grant agreements No. 690575 and 674896. 
\end{acknowledgements}

\bibliographystyle{spphys}       
\bibliography{refs}


\end{document}